\documentclass[
    prd,
    noeprint,
    nofootinbib,
    reprint,
    superscriptaddress,
]{revtex4-2}

\usepackage{microtype}
\usepackage[varl]{zi4}
\usepackage{tabularray}

\usepackage{verbatim}

\usepackage{mathtools,amssymb}

\usepackage[dvipsnames,svgnames]{xcolor}
\usepackage[
    colorlinks,
    linkcolor=RoyalBlue!80!Black,
    citecolor=RoyalBlue!80!Black,
    urlcolor=ForestGreen!60!Black,
]{hyperref}
\usepackage[capitalise]{cleveref}
\usepackage[
  range-phrase=--,
  range-units=single,
]{siunitx}

\usepackage{tikz}
\usetikzlibrary{shapes.arrows,arrows.meta,fadings}

\usepackage{graphicx}
\graphicspath{{figures/}}
\usepackage{orcidlink}

\newcommand{\rmd}{\mathrm{d}}
\newcommand{\rme}{\mathrm{e}}
\newcommand{\rmi}{\mathrm{i}}
\newcommand{\eps}{\varepsilon}

\DeclareMathOperator{\erf}{erf}
\DeclareMathOperator{\real}{Re}
\DeclareMathOperator{\imag}{Im}

\newcommand{\prlsection}[1]{\vspace*{\baselineskip}\noindent\emph{#1 ---}}

\newcommand{\edh}[1]{{\color{Teal}\small\textsf{#1}}}

\begin{document}

\title{Coating-free monolithic fused silica resonator via total internal reflection for precision laser stabilization}
\author{Swadha Pandey\,\orcidlink{0000-0002-2426-6781}}
\affiliation{LIGO Laboratory, Department of Physics, Massachusetts Institute of Technology, Cambridge, MA 02139, USA}
\author{Rodica Martin}
\affiliation{Montclair State University, Montclair, NJ 07043, USA}
\author{Peter Fritschel}
\affiliation{LIGO Laboratory, Department of Physics, Massachusetts Institute of Technology, Cambridge, MA 02139, USA}
\author{Matthew Evans\,\orcidlink{0000-0001-8459-4499}}
\affiliation{LIGO Laboratory, Department of Physics, Massachusetts Institute of Technology, Cambridge, MA 02139, USA}
\author{Evan D. Hall\,\orcidlink{0000-0001-9018-666X}}
\email{evanhall@mit.edu}
\affiliation{LIGO Laboratory, Department of Physics, Massachusetts Institute of Technology, Cambridge, MA 02139, USA}

\date{\today}

\begin{abstract}
Brownian noise in the thin-film mirror coatings of optical reference cavities is a fundamental limitation in precision metrology, including gravitational-wave detectors and optical atomic clocks. 
We demonstrate a monolithic fused silica resonator that eliminates the use of thin-film coatings by operating via total internal reflection (TIR), achieving a finesse of 1225 and an optical mode volume of \qty{50}{mm^3}. 
To our knowledge, this is the largest mode volume reported for any coating-free monolithic resonator, comparable to state-of-the-art reference cavities. 
We use the cavity to frequency-stabilize an Nd:YAG laser, and demonstrate its application to precision metrology by operating it in a passive ring gyroscope configuration. 
This platform circumvents the dominant noise source of conventional reference cavities and provides a pathway toward cryogenic silicon TIR resonators that could surpass current frequency stabilization limits.
\end{abstract}

\maketitle

Optical resonators enable the production and manipulation of highly coherent, phase-stable light, and are foundational across precision measurement.
They are used for frequency stabilization in gravitational-wave interferometry on the ground~\cite{Kwee:2012usw,Adhikari:2013kya} and in space~\cite{LISAfreq,TianQin:2015yph}, for gravimetry~\cite{Abich:2019cci}, and for optical atomic clocks~\cite{Ludlow:2015gnc}.
Frequency stabilization in these cavities is achieved by referencing the laser light to a passive, rigid optical resonator, commonly using rf phase modulation~\cite{Drever:1983qsr}.
When this scheme is applied to a cavity with an optical mode at frequency $\nu$ and round-trip propagation length $L$, fractional path length fluctuations $\Delta L / L$ within the reference cavity are imprinted onto fractional frequency fluctuations $\Delta\nu/\nu$ of the stabilized laser.
Many of the fundamental mechanisms contributing to path length fluctuations\,---\,in particular, thermodynamically driven fluctuations in the resonator's shape and temperature\,---\,favor a large optical mode volume, realizable with a free-space cavity using discrete Bragg mirrors bonded to a hollow spacer. 
The stability of reference cavities, such as those used in interferometric gravitational-wave detectors and optical atomic clocks, is currently limited by the Brownian motion of the Bragg mirror coatings.
These coatings are often fabricated from pairs of metal oxides, such as silica and titania-doped tantala~\cite{Granata:2019idz}, and a significant effort is currently directed toward developing Bragg stacks with superior thermal noise performance that can also satisfy stringent demands on optical absorption and scatter~\cite{Craig:2019,Tait:2020}.
These superior stacks may be formed from alternate metal oxides, which can offer several times less Brownian noise than the current state of the art~\cite{Vajente:2021,Amato:2021ryz,McGhee:2023lpm}.
Semiconductor stacks such as GaAs/AlGaAs offer an even greater Brownian noise reduction, but introduce additional intrinsic noises that may exceed the Brownian noise limit~\cite{Cole:2013,Kedar:2022alu,Yu:2022mbe,Lee:2025mbd}.

Monolithic optical resonators\,---\,wherein a single material defines both the resonator volume and the optical mode boundaries\,---\,do not require coatings or other specially engineered surface structures, and find wide application in physics and metrology.
Platforms include microring resonators~\cite{Bogaerts:2012nzf}, spherical or toroidal resonators supporting whispering-gallery modes~\cite{2006IJSTQ..12....3M,2006IJSTQ..12...15I,2007JOSAB..24.1324M}, and prism resonators that confine a paraxial Gaussian mode through a discrete number of total internal reflections (TIR).
This last category of resonator is particularly well-suited to support large mode volumes, limited mainly by the size of available substrates and the fabrication tolerance of the faces.
Experiments have demonstrated evanescent coupling of laser light into millimeter- and centimeter-scale cavities~\cite{1992OptL...17..378S}, including for active applications such as parametric amplification~\cite{1993JOSAB..10.1696S,Brieussel:2016zpa}.

In this \emph{Letter}, we demonstrate a monolithic total internal reflection resonator (MOTIRR) in fused silica, operating at \qty{1064}{\nm} without the use of thin-film coatings.
We have used the resonator to frequency-stabilize an Nd:YAG NPRO laser, and additionally have operated the resonator in a Sagnac gyroscope configuration.
To our knowledge, the optical mode volume of our resonator\,---\,\qty{50}{\mm^3}\,---\,is the largest so far achieved in any monolithic, coating-free optical resonator, and is comparable to state-of-the-art rigid optical reference cavities.
The large mode volume suppresses thermodynamic noise, making this a promising platform for demanding laser frequency stabilization applications, with a natural upgrade path to cryogenic silicon for performance beyond current limits.

\prlsection{Design and fabrication}
\begin{figure*}[t]
    \centering
    \includegraphics[width=\textwidth]{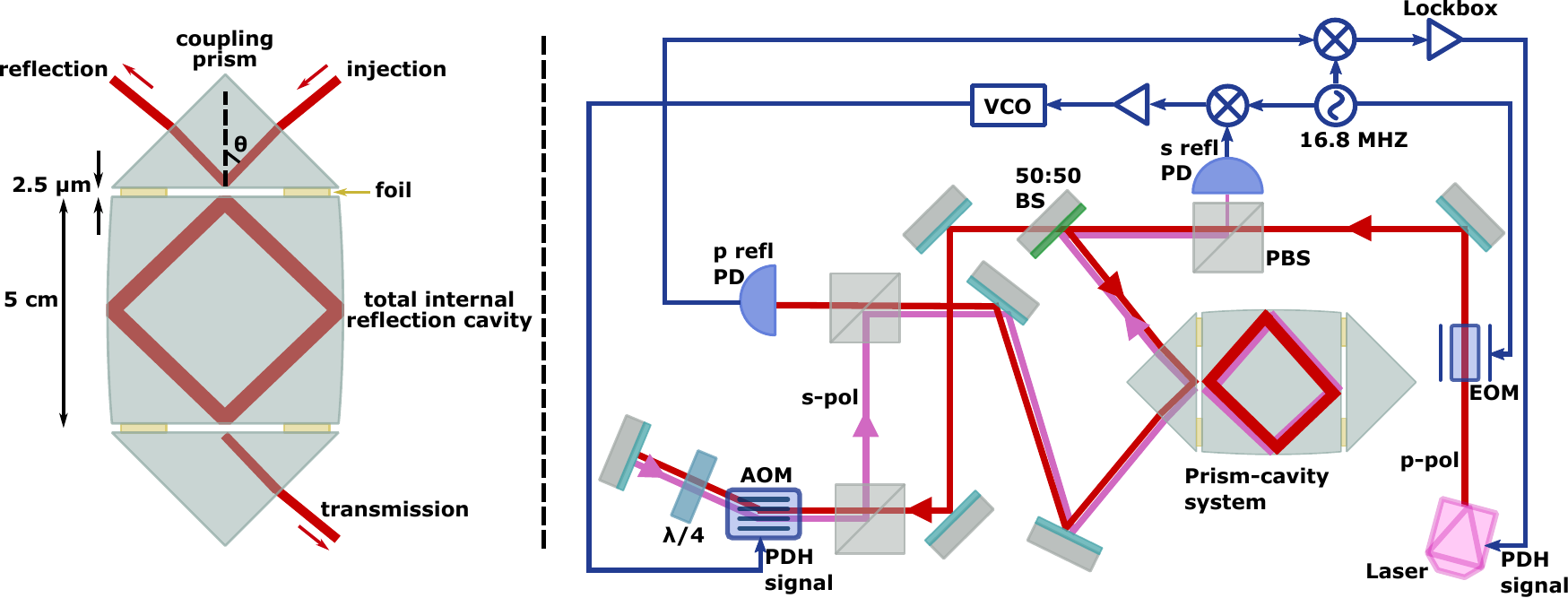}
    \caption{\emph{Left:} diagram of optical cavity formed from a $\qty{5.08}{cm} \times \qty{5.08}{cm}$ fused silica cuboid with two flat and two curved faces. Light is coupled in and out of the cavity via frustrated total internal reflection through two triangular prisms spaced from the cavity by \qty{2.5}{\micro\meter} thick aluminum foil. \emph{Right:} optical layout of the experiment, where p-polarized light from a \qty{1064}{\nm} laser is coupled into the cavity and resonates in the counter-clockwise direction. The same laser light is converted to s-polarization and sent through a double-pass acousto-optic modulator (AOM) to match the Goos--Hänchen separation and then coupled back into the cavity to resonate in the clockwise direction. Both polarizations are locked to the cavity using the Pound--Drever--Hall (PDF) technique. This setup is used to measure the freerunning noise in the system as well as make a differential measurement using two counterpropagating TEM$_{(0,0)}$ modes in orthogonal polarizations.}
    \label{fig:layout}
\end{figure*}
\begin{figure}[t]
    \centering
    \includegraphics[width=\columnwidth]{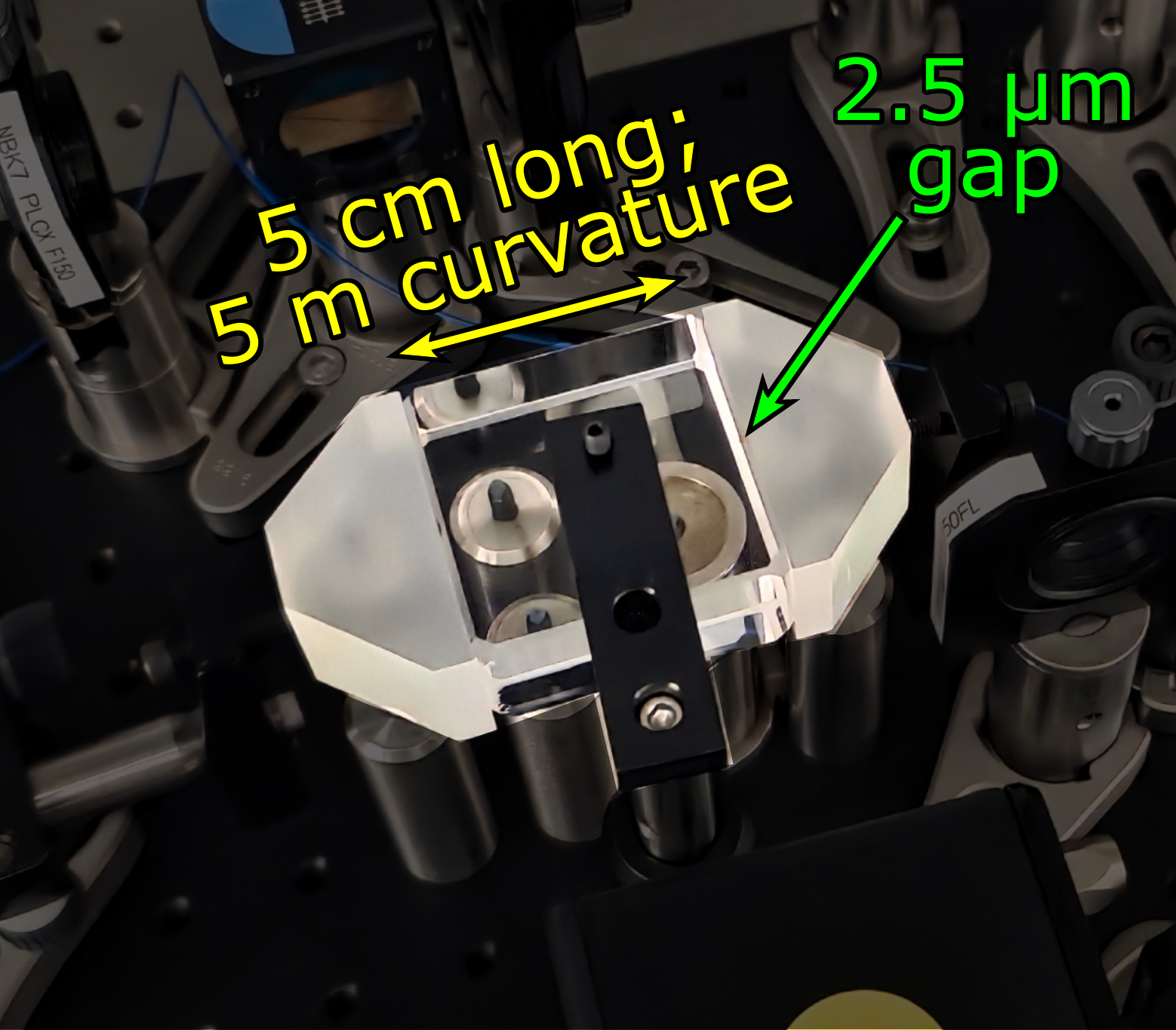}
    \caption{Photograph of the fused silica optical cavity described in \cref{fig:layout}. The prism--cavity distance is set using aluminum foil spacers and the setup is clamped together.}
    \label{fig:picture}
\end{figure}
The optical cavity design is shown in \cref{fig:layout}, consisting of a $\qty{5.08}{cm} \times \qty{5.08}{cm} \times \qty{2.54}{cm}$ rectangular prism of IR-grade fused silica.
Of the four $\qty{5.08}{\cm}\times\qty{2.54}{\cm}$ faces, two of the opposing faces are flat and the other two opposing faces are convex with a \qty{5.0}{\meter} radius of curvature.
These four faces are capable of supporting Gaussian traveling wave modes by total internal reflection, with a nominally square path with optical length $n L = 4n\times\sqrt{2}\times\qty{2.54}{\cm} = \qty{20.7}{\cm}$, where $n = 1.45$ is the index of refraction of fused silica at \qty{1064}{\nm}.
The mode waist is \qty{320}{\micro\meter} in the transverse plane and \qty{380}{\micro\meter} in the saggital plane, and the spot size variation over the optical path length is small due to the cavity's near-unity stability factor.
Adjacent faces of the cavity are toleranced to be perpendicular to each other to within \qty{0.1}{\degree}, to ensure the cavity mode forms within \qty{1}{\cm} of the center of each face.
This was set using cavity stability analysis and verified with numerical simulations that show that the  cavity mode falls within this specified region \qty{85}{\%} of the time (see Supplementary Information).

Light is coupled in and out of the cavity by slightly frustrating the internal reflection at the two flat faces using two triangular prisms, where the power coupling level is set by the distance between each prism and the cavity face.
This distance is fixed using three pieces of \qty{2.5}{\micro\meter} thick aluminum foil to separate each prism face from its corresponding cavity face.
Nominally, the \qty{45}{\degree} angle of incidence implies power transmissivities of \qty{0.09}{\%} for an s-polarized mode, and \qty{0.3}{\%} for a p-polarized mode~\cite{1992OptL...17..378S}.
%
%
Due to fabrication tolerances, the cavity mode is found to form off-center on the various faces, and the angles of incidence on these faces, and hence the transmissivities, may deviate from the design value.

\prlsection{Cavity metrology}
Being made of low-OH-content fused silica, the cavity substrate is expected to have an absorption $\lesssim\qty{1}{\dB/\km}$, which is near the limit set by Rayleigh scattering and is negligible in this application~\cite{Jackson:1998nia}.
More relevant is the microroughness of the cavity's surfaces, which we determined by optical profilometry to have rms roughness in the range \qtyrange{1.0}{1.3}{\nm}.
In the limit that this microroughness simply scatters light out of the fundamental Gaussian mode without any subsequent coherent effects, this would imply a total integrated scatter of \qty{400}{ppm}, accounting for the non-normal angle of incidence~\cite{1983ApOpt..22.3207E}.
On the other hand, when the cavity's TEM$_{(0,0)}$ mode is pumped with laser light, we find that some power is transferred to the partially resonant TEM$_{(0,22)}$ mode, which is located at a \qty{2}{\MHz} detuning from the cavity fundamental; a fractional power coupling of about \qty{1}{ppm} per bounce is needed between these two modes to explain the observed excitation of the higher order mode.
We find a finesse of 795 and 1225 for the TEM$_{(0,0)}$ modes of the p and s polarizations, respectively.
These finesses are smaller than the expectation from the combination of the theoretical transmittance values ($T_p$, $T_s$) and the budgeted \qty{400}{ppm} of surface scatter, and could indicate that the internal optical loss from scatter or absorption is closer to \qty{0.3}{\%} per round trip.
The p and s modes are separated by a detuning of \qty{391.6}{\MHz}.
The theoretical expectation from the Goos--Hänchen effect, including any contributions from the slight frustration of the internal reflection, is \qty{397}{\MHz}~\cite{1992OptL...17..378S}.
The deviation is possibly explained by additional perturbations to the TEM$_{(0,0)}$ modes' axial eigenfrequencies due to coupling with other modes.

\prlsection{Laser stabilization}
\begin{figure}[t]
    \centering
    \includegraphics[width=\columnwidth]{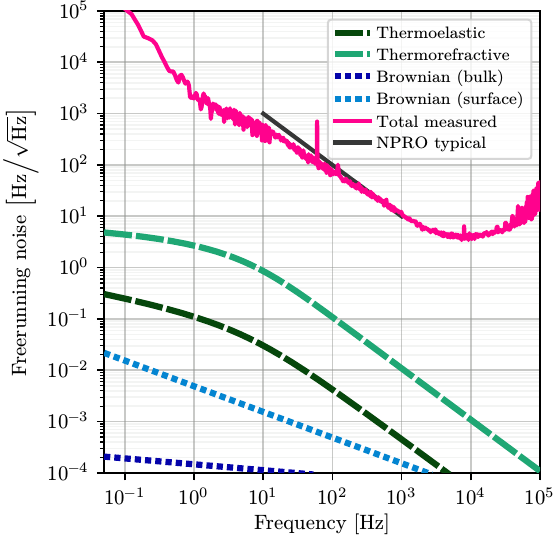}
    \caption{
      Measured amplitude spectral density of the freerunning noise of the laser relative to the cavity's p-polarized TEM$_{(0,0)}$ mode. The measurement from \qtyrange{10}{1000}{Hz} is consistent with the performance of Nd:YAG NPROs.
      The fundamental cavity displacement noises limiting the system are expected to be thermorefractive, thermoelastic, and Brownian noises.
      Relevant cavity parameters are given in the text and supplement.
    }
    \label{fig:noise}
\end{figure}
We stabilized a commercial \qty{1064}{\nm} Nd:YAG laser to the cavity's TEM$_{(0,0)}$ mode in the p polarization by applying \qty{16.8}{\MHz} sidebands to derive a Pound--Drever--Hall error signal in reflection, and filtering it to produce a frequency feedback signal.
\cref{fig:noise} shows the spectral density of the freerunning noise in the system, calculated from the loop-corrected control signal.
The magnitude and spectral shape of the measured noise from \qtyrange{10}{1000}{\Hz} are typical of such lasers, arising from thermal distortions in the gain medium driven by fluctuations of the laser pump light; the upturn above \qty{10}{\kHz} is consistent with relaxation oscillation~\cite{Vahlbruch:2025tmh}.

The fundamental cavity noises, also shown in \cref{fig:noise}, are expected to lie several orders of magnitude below the laser noise, as shown in \cref{fig:noise}. 
Chief among these is thermorefractive noise~\cite{2011PhRvA..84a1804A,2019PhRvA..99f1801H,Panuski:2020cxc}\,---\,refractive index fluctuations due to thermodynamically driven temperature fluctuations\,---\,calculated here using established methods for Gaussian beams~\cite{2004PhLA..324..345B,2008PhLA..372.1941L}.
For light propagating through a physical path length $L$ in a material of density $\rho$, specific heat $C$, thermal conductivity $\kappa$, and thermorefractive coefficient $\partial n/\partial T$, the power spectral density of fractional frequency fluctuations $S_{\nu\nu}^{\text{(TR)}}(\Omega)/\nu^2$ due to thermorefractive noise has a characteristic value $(\partial n/\partial T)^2 k_\text{B} T^2 / \pi\kappa L$ at a Fourier frequency $\Omega_w = 4\kappa/\rho C w^2$, which corresponds to the relaxation time of thermal fluctuations across the transverse mode size $w$.
The spectral density scales as $(\Omega_w/\Omega)^2$ above this frequency, and as $\ln(\Omega_w / \Omega)$ below it.
The same temperature fluctuations also drive thermoelastic deformations, although in fused silica the effect of these deformations on the optical path length is subdominant to the refractive index fluctuations because the coefficient of thermal expansion is small ($\lesssim\qty{e-6}{\kelvin^{-1}}$) compared to the thermorefractive coefficient (\qty{9e-6}{\kelvin^{-1}}).

Brownian motion of the cavity surfaces due to internal friction also contribute to optical path length fluctuations.
Audio-band vibration measurements indicate that in high quality fused silica samples, the bulk dissipation $1/Q_\text{bulk}$ can be nearly viscous and exceed \num{e9} near \qty{100}{\Hz}~\cite{Cumming:2012zz}, whereas the dissipation near the surface is much higher, and depends on damage from abrasive polishing and subsequent surface treatments~\cite{1998RScI...69.3681S}.
In flame- or laser-polished samples, the product of the surface layer thickness $d_\text{surf}$ and the layer's dissipation $1/Q_\text{surf}$ has been reported to be as low as \qty{6}{\pm}~\cite{Cumming:2012zz}; our mechanically polished cavity faces are unlikely to achieve this.
Nonetheless, we expect that with appropriate surface treatments, it would be able to outperform Bragg reflector coatings, where a coating with thickness $d_\text{coat}$ of a few microns and a state-of-the art dissipation $1/Q_\text{coat} \approx \num{e-5}$ would still be limited to a depth--loss product of tens of picometers or more.
A benchmark value of $d_\text{surf}/Q_\text{surf}$ is provided by the quotient $w/Q_\text{bulk}$; when the latter exceeds the former, the surface contributions to the cavity Brownian noise become subdominant to the contributions from the bulk~\cite{Nakagawa:2001di}.
A monolithic, coating-free cavity offers the prospect that with a sufficiently large beam size $w$ and with proper surface treatments, this can be achieved for materials with high $Q_\text{bulk}$.

\begin{figure}[t]
    \centering
    \includegraphics[width=\columnwidth]{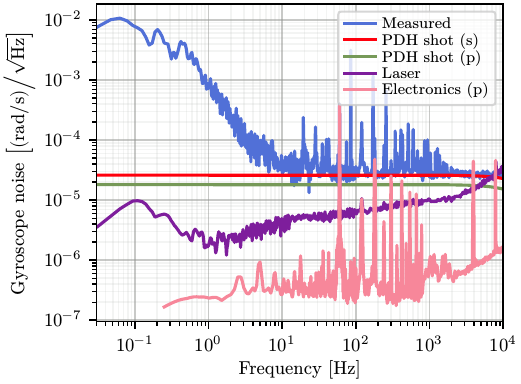}
    \caption{
      Measured amplitude spectral density of the observed differential frequency noise between counterpropagating p- and s-polarized TEM$_{(0,0)}$ modes, calibrated into equivalent cavity rotation (blue).
      It is found to be limited by the shot noise on the s-polarization reflection photodiode (red) above \qty{10}{\Hz}.
      Shot noise on the p-polarization reflection photodiode (green), residual Nd:YAG NPRO laser noise (purple), and input-referred noise from the p-polarization control electronics (pink) are also shown.
    }
    \label{fig:gyro_noise}
\end{figure}

\prlsection{Application: gyroscope}
To demonstrate the feasibility of our cavity for metrology applications, we operated it in a passive ring gyroscope configuration.
We pumped counterpropagating TEM$_{(0,0)}$ modes of the same axial order; the optical diagram is shown on the right side of \cref{fig:layout}.
Light from our single Nd:YAG laser is split into p and s polarized light; the laser is stabilized to the cavity by resonating the p-polarized light traveling in the counterclockwise direction and sensing its PDH error signal in reflection.
Separately, the s polarized light is doubly passed through an acousto-optic modulator (AOM) to achieve a \qty{391.6}{\MHz} frequency shift to match the Goos--Hänchen separation; this light is made to resonate in the clockwise direction in the cavity by sensing its PDH signal in reflection and feeding back to the frequency control applied to the AOM.
Within the bandwidth of the p and s feedback loops, this control signal tracks the differential frequency fluctuation $\Delta\nu_{\text{ps}}$ between the two counterpropagating beams.
Via the Sagnac effect, this differential frequency fluctuation can be referred to an equivalent rotational velocity:
the physical path $L$ taken by either the p or s light defines a directed area $\mathbf{A}$, so that if the cavity has an angular velocity $\boldsymbol{\Omega}$, the effect induces a differential frequency shift $(4 /\lambda_0 L n) \mathbf{A} \cdot \boldsymbol{\Omega}$, with $\lambda_0$ being the vacuum wavelength of the light~\cite{Post:1967qwl,Chow:1985zz,2006OptCo.259..393M}. 
\cref{fig:gyro_noise} shows the observed frequency splitting, calibrated by the Sagnac factor into an equivalent rotation of the cavity.
Down to \qty{10}{\Hz}, the observed noise is consistent with the limit imposed by the shot noise of the fraction of the carrier light not coupled into the cavity's transverse spatial eigenmode; there are two such contributions from the two frequency discriminant readouts of the counterpropagating modes.
(We observed that on resonance, \qty{51}{\%} of the incident p polarized light power was coupled into the corresponding TEM$_{(0,0)}$ mode, while for the s polarization the fraction was \qty{32}{\%}.)
The increased noise below \qty{10}{\Hz} is not correlated with ground vibration and may be due to temperature or other environmental fluctuations in the vicinity of the acousto-optic modulator, which is outside the vacuum system and differentially affects the counterpropagating beams.

In this configuration, residual frequency noise from the laser is largely common to the two loops, and is therefore significantly rejected in the differential signal, as shown in \cref{fig:gyro_noise}. 
We also expect the thermorefractive, thermoelastic, and Brownian noises to cancel due to common mode rejection, meaning that, if technical noise sources are sufficiently controlled, the gyroscope configuration will be limited by shot noise.
The MOTIRR's applicability for gyroscopy is enhanced by the fact that it presents a monolithic, centimeter-scale construction with a mode volume still large enough to avoid significant nonlinear effects~\cite{2018PhLA..382.2289M}.
In this respect, it occupies a middle ground between passive gyroscopes in a WGM form factor with generally smaller mode volume (e.g., Ref.~\cite{2017Optic...4..114L}), and larger free-space ring cavities with multiple mirrors (e.g., Refs.~\cite{1977ApPhL..30..478E,Korth:2015nsa,Martynov:2018ljn,Liu:2019ewx,Zhang:2020gux,2025OptL...50.3883C,2026MeScT..37y5203G,2013RScI...84d1101S}).

\prlsection{Conclusion}
We have demonstrated a monolithic fused silica resonator that supports a \qty{50}{\mm^3} Gaussian mode with finesse exceeding \num{e3}, operating by total internal reflection and without the use of thin-film coatings.
We have used it to stabilize an Nd:YAG laser, and demonstrated its application to precision metrology by showing that it can be used as a passive ring gyroscope.

We anticipate a clear upgrade pathway to turn the centimeter-scale MOTIRR demonstrated here into a more sensitive frequency reference, either in relative terms (for gyroscopic or similar measurements) or in absolute terms, for laser stabilization applications.
An improvement in the relative frequency stability of the cavity can be made with increased cavity finesse, using a larger gap size between the main cavity prism and the coupling prisms, using more optical power for the counterpropagating modes, and critically coupling the optical modes to the cavity, in addition to technical improvements such as in-vacuum acousto-optic modulation.
A finesse $\mathcal{F} \sim \num{e4}$ and input power $P = \qty{1}{\mW}$ offers the prospect of a shot-noise-limited relative frequency stability of ${\sim}\qty{1}{\mHz/\sqrt{\Hz}}$ for counterpropagating modes, and hence a rotational sensitivity better than \qty{100}{(\nano\radian/\second)/\sqrt{\Hz}} for our cavity geometry, or an angular random walk of \qty{e-4}{\degree/\sqrt{\unit{\hour}}}, which would offer navigation-grade performance~\cite{2023ISenJ..2329948D}.
Attention must be paid to differential noise sources; for instance, if acousto-optic modulation is retained as the means to shift light between the clockwise and counterclockwise modes, the frequency noise of the rf oscillator must also be kept smaller than the desired relative frequency stability of the overall sensing scheme.

Applications requiring absolute frequency stability, or relative-frequency applications where the common-mode cancellation of noises in different optical modes is imperfect, motivate further improvements to MOTIRRs to reduce fundamental and technical noise sources.
Good fundamental noise performance requires attention to control of Brownian noise due to the reflective cavity surfaces, the mounting or bonding scheme used to fix the coupling prisms relative to the cavity, and the overall mounting of the cavity inside the vacuum chamber.
For environmentally driven vibrations, a simple elastic model shows that our MOTIRR couples vertical acceleration to cavity path length fluctuation with a typical fractional susceptibility $\qty{e-10}{\vphantom{1}/(\meter/\second^2)}$ (see Supplement), although as with free-space cavities, it may be possible to find a geometry of the cavity or its mount that has a reduced vibrational susceptibility~\cite{2012NaPho...6..687K}.
In the most demanding applications, active vibrational isolation with inertial sensors may be required.
Environmental temperature fluctuations must also be controlled: a benchmark value for such fluctuations is set by the power spectral density of thermodynamically driven temperature fluctuations at low frequency, $k_\text{B} T^2 / \pi \kappa L$~\cite{2015OExpr..23.5134D}, with $T$ being the nominal cavity temperature and $\kappa$ the thermal conductivity of the bulk.
The cavity temperature will also fluctuate due to absorbed beam power~\cite{Braginsky:1999rp}; in the limit of weak thermal coupling to the environment, the shot noise from an absorbed power $P_\text{abs}$ causes a temperature fluctuation with amplitude spectral density $\sqrt{S_{TT}(\Omega)} = \left.\sqrt{2 h \nu P_\text{abs}} \middle/(\rho C V \Omega)\right.$ on timescales longer than the thermal relaxation time across the cavity volume $V$.
For our room-temperature silica cavity, the resulting requirements on absorbed power are not so stringent: with $T \sim \qty{300}{\kelvin}$ and $\kappa = \qty{1.4}{\watt/(\kelvin\,\meter)}$, the thermodynamic fluctuations are of order \qty{1}{\nano\kelvin/\sqrt{\Hz}}, while the photothermal shot noise fluctuations even for \qty{100}{\micro\watt} of absorbed power are \qty{10}{\pico\kelvin/\sqrt{\Hz}} at \qty{1}{\mHz}.

In the longer term, greater stability could be achieved with a cryogenic resonator where thermodynamic fluctuations are lower than at room temperature, and for some materials the optical and mechanical susceptibilities to such fluctuations are greatly reduced.
A silicon resonator operating at \qty{20}{\kelvin}, for instance, has zero thermal expansion, a thermorefractive coefficient $\partial n/\partial T \simeq \qty[retain-unity-mantissa=true]{e-6}{\kelvin^{-1}}$, and a conductivity exceeding \qty{e3}{\watt/(\kelvin\,\meter)}~\cite{2012ApPhL.101d1905K,2015PhRvB..92q4113M}.
In combination with appropriate treatments to minimize mechanical surface loss~\cite{Nawrodt:2013uca}, a macroscopic, high-purity silicon resonator offers the possibility of greatly reduced thermal noise compared to similarly sized free-space resonators based on Bragg stacks, and hence greatly increased absolute frequency stability, perhaps approaching \qty{1}{\mHz/\sqrt{\Hz}} on millihertz timescales (see Supplementary Information).
Achieving such stability requires even more stringent control of vibration, photothermal fluctuations, temperature stability, and other technical issues such as rf amplitude modulation~\cite{Kedar:2023olv}.

\prlsection{Acknowledgments}
The authors thank the MIT Kavli Institute for supporting this work.

\bibliography{references}

\appendix

\section{Mode structure}

\begin{figure}[t]
    \centering
    \includegraphics[width=\columnwidth]{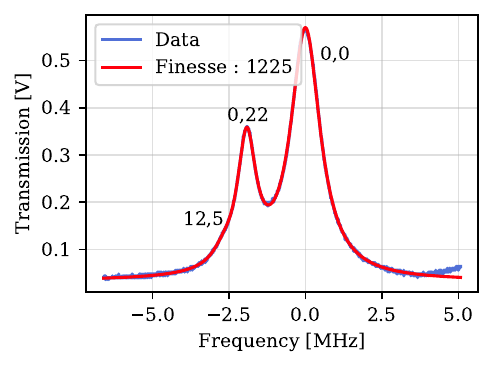}
    \caption{Cavity mode structure for s-polarized light: measured transmission (blue) and Lorentzian fit (red) near the TEM$_{(0,0)}$ mode, with a cavity finesse of 1225. Also visible are resonantly enhanced higher-order modes TEM$_{(0,22)}$ and TEM$_{(12,5)}$, coupled to TEM$_{(0,0)}$ via scattering due to fabrication irregularities.}
    \label{fig:finesse}
\end{figure}

\begin{figure}[t]
    \centering
    \includegraphics[width=\columnwidth]{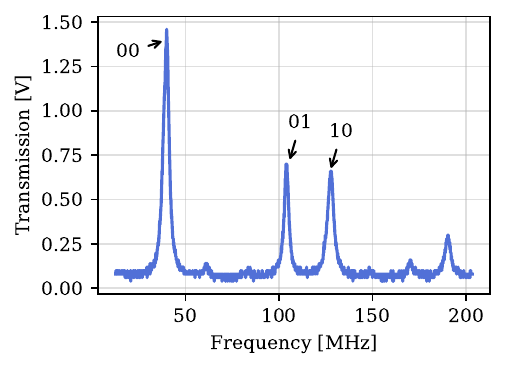}
    \caption{Cavity mode structure for p-polarized light, with transmission data showing the mode spacing between the TEM$_{(0,0)}$ mode and higher-order modes including TEM$_{(0,1)}$ and TEM$_{(1,0)}$.}
    \label{fig:mode_structure}
\end{figure}

To measure the transverse spatial mode structure of our cavity, we drive the NPRO laser's calibrated piezoelectric transducer with a triangular wave to scan the cavity length and obtain transmission spectra for both polarizations.
\cref{fig:finesse} shows the mode structure for s-polarized light, where higher-order modes TEM$_{(0,22)}$ and TEM$_{(12,5)}$ are resonantly enhanced via scattering from the TEM$_{(0,0)}$ mode caused by fabrication irregularities in the glass cavity. The cavity finesse for s-polarization is measured to be 1225. \cref{fig:mode_structure} shows the mode spacing for p-polarized light, highlighting the TEM$_{(0,0)}$ mode and higher-order modes including TEM$_{(0,1)}$ and TEM$_{(1,0)}$.

\section{Surface roughness}

\begin{figure}[t]
    \centering
    \includegraphics[width=\columnwidth]{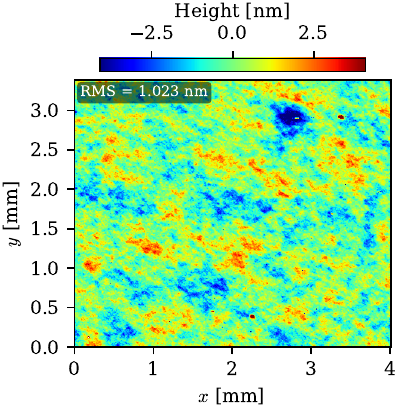}
    \caption{Measured surface roughness of one of the curved sides of the fabricated glass cavity, where the rms value was found to be  \qty{1.0}{nm}.}
    \label{fig:roughness}
\end{figure}

We measured the microroughness of each surface of the fabricated glass cavity using the 4D Technology NanoCam optical profiler. We found this rms roughness to be in the range \qtyrange{1.0}{1.3}{\nm}. The measured surface roughness for one of the curved cavity surfaces is shown in \cref{fig:roughness}, for which the rms roughness was found to be \qty{1.0}{\nm} over a measured area of \qty{4.26}{\mm}$\times$\qty{3.55}{\mm} in the central region of the surface.

\section{Tolerancing and cavity mode}

For a two-mirror cavity with round-trip length $L$ and concave mirrors of equal radius of curvature $R_1 = R_2 = R$, the coupling between mirror misalignment $\theta_1 = \theta_2 = \theta$ and beam spot displacement $\Delta x$ is given by~\cite{Siegman:1986}
\begin{equation}
    \Delta x = \frac{1}{1-g}\times\frac{L\theta}{2}, \qquad g = 1 - \frac{L}{2R}.
\end{equation}
For our fabricated cavity of dimensions $\qty{5.08}{\cm} \times \qty{2.54}{\cm}$ on the faces, we require the cavity mode to form within the region of high optical finish, setting a maximum allowed beam spot displacement of $\Delta x_\mathrm{max} = \qty{10}{\mm}$. This gives a misalignment tolerance of $\theta_\mathrm{max} = \qty{6.9}{arcmin}$ for $L = \qty{14.14}{\cm}$ and $R = \qty{5}{\m}$, which we take as an estimate for the more complex four-mirror ring cavity geometry.

\begin{figure*}
    \centering
    \includegraphics[width=\textwidth]{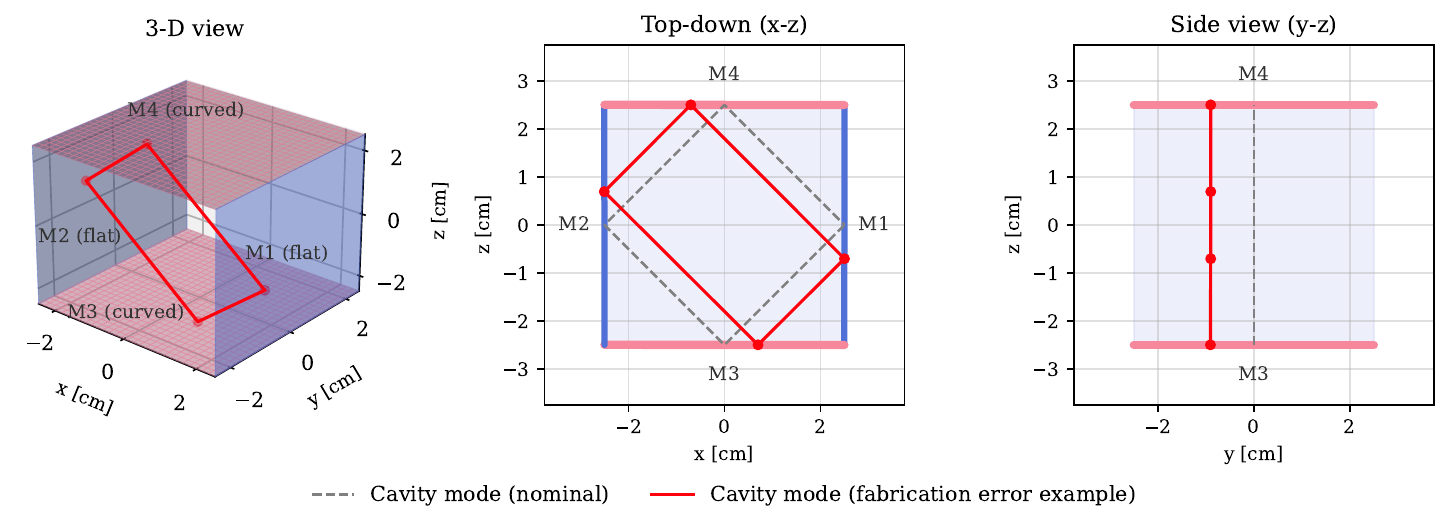}
    \caption{Visualization of the cavity mode deviation from the nominal symmetric mode due to fabrication errors of up to \qty{6}{arcmin} in pitch and yaw and \qty{1}{\mm} in center-of-curvature position at each surface. Each glass surface acts as a mirror via total internal reflection (TIR); M1 and M2 denote the flat sides, and M3 and M4 the curved sides of radius \qty{5}{\m}. \emph{Left:} Three-dimensional view of the cavity mode for a randomly drawn set of fabrication errors. \emph{Middle:} X-Z projection of the perturbed (red) and nominal (gray) modes. \emph{Right:} Y-Z projection of the perturbed (red) and nominal (gray) modes.}
    \label{fig:geometry}
\end{figure*}

\begin{figure}[t]
    \centering
    \includegraphics[width=\columnwidth]{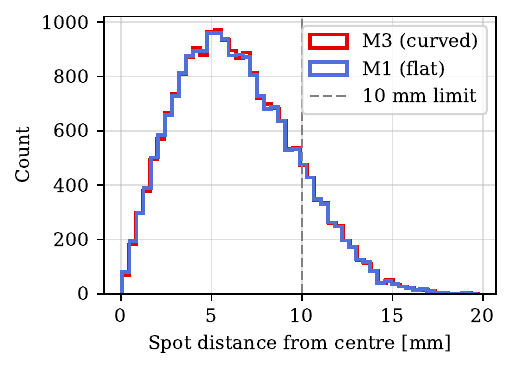}
    \caption{Histogram of beam spot distance from center at a flat surface, M1 (blue), and a curved surface, M3 (red), from 20,000 simulations of fabrication errors drawn uniformly within \qty{6}{arcmin} in pitch and yaw at each surface and \qty{1}{\mm} in center-of-curvature position at each curved surface. The vertical gray line marks the \qty{10}{\mm} threshold for acceptable mode formation. For these tolerances, the beam spot falls within the acceptable region \qty{85}{\%} of the time.}
    \label{fig:modpos}
\end{figure}

To verify this tolerance for our design, we numerically simulate the 
three-dimensional cavity geometry under fabrication errors. We model 
the cavity as a cube of side \qty{5}{\cm} for simplicity. The fabricated cavity has mirror face dimensions of \qty{5}{\cm} $\times$ \qty{2.5}{\cm}, but since our beam spot threshold of \qty{10}{\mm} is well within both transverse dimensions, this approximation does not affect the results. Each surface is assigned independent pitch and yaw misalignments drawn uniformly within \qty{6}{arcmin}, and each curved surface is additionally assigned a center-of-curvature displacement drawn uniformly within \qty{1}{\mm}. For each set of drawn errors, the cavity mode is found by solving the round-trip ray propagation through the perturbed geometry. An example of the resulting mode deviation from the nominal symmetric mode is shown in \cref{fig:geometry}.

We repeat this procedure for 20,000 independent draws to produce the 
distributions of beam spot position shown in \cref{fig:modpos}. The 
distributions at the flat surface M1 and curved surface M3 are found to be identical, and \qty{85}{\%} of simulations yield a beam spot within the \qty{10}{\mm} threshold at both surfaces, confirming that the specified fabrication tolerances are sufficient for reliable mode formation. 

\section{Noise terms}

We consider in more detail the basic noise terms that we expect to contribute to the overall stability of a monolithic TIR resonator.
Two cases, a room-temperature fused silica cavity and a cryogenic silicon cavity, are considered, and their parameters are given in \cref{tab:cavity parameters}.
In both cases the nominal angle of incidence is \qty{45}{\degree}, and the mode $g$ factor is very close to unity, so that the cavity mode is of nearly uniform elliptical cross-section, with transverse and sagittal waists $w_\text{t}$ and $w_\text{s}$; below, any appearance of $w$ is taken to be the geometric mean $\sqrt{w_\text{t} w_\text{s}}$ unless otherwise specified.

An ideal Pound--Drever--Hall frequency locking scheme tracks variations in the phase accumulated by the laser light, of optical frequency $\nu$, propagating around the cavity.
This phase is $\Phi = 2\pi\nu n L/c$, where $n$ is the index of refraction of the cavity medium, $L$ is the physical round-trip path length, and $c$ is the vacuum speed of light.
Correspondingly, deviations in the accumulated phase may arise from a frequency detuning $\delta\nu$, a refractive index shift $\delta n$, or a physical path length perturbation $\delta L$:
\begin{equation}
  \delta\Phi = (2\pi/c)(\delta\nu\,n\,L + \nu\,\delta n\,L + \nu\,n\,\delta L).
  \label{eq:dPhi PDH}
\end{equation}
Since an ideal frequency locking scheme suppresses this phase deviation, fluctuations in the refractive index or path length are impressed onto the laser light, and can therefore be expressed as an equivalent laser frequency fluctuation.

The next subsections give noise estimates for cavity Brownian noise, thermorefractive noise, thermoelastic noise, vibrational noise, shot noise, residual laser frequency noise, and photothermal noise.
All spectral densities given are single-sided.

\subsection{Brownian noise}

We consider Brownian noise from two sources: thermally driven volume fluctuations in the bulk of the cavity, and fluctuations arising from a thin and more lossy surface layer.
Taking the beam diameter $2w$ to be much smaller than the cavity substrate dimensions $D$ and $H$, one can use the result previously derived for a Bragg-coated mirror~\cite{Nakagawa:2001di,Harry:2001iw}:
\begin{equation}
  S_{xx}(\Omega) = \frac{4k_\text{B}T(1-\sigma^2)}{\sqrt{\pi} w E \Omega} \left[ \phi_\text{bulk} + \frac{2}{\sqrt{\pi}}\left(\frac{1-2\sigma}{1-\sigma}\right) \frac{d_\text{surf}\phi_\text{surf}}{w}\right],
\end{equation}
where $E$ is the Young modulus, $\sigma$ the Poisson ratio, and $T$ the temperature.
The bulk loss angle $\phi_\text{bulk}$ and the loss-depth product $d_\text{surf} \phi_\text{surf}$ for the surface contribution are taken from previous measurements described in in \cref{tab:cavity parameters}.
The assumption $d_\text{surf} \phi_\text{surf} = \qty{6}{\pm}$ is likely optimistic for our current cavity, as the surfaces were not flame- or laser-polished, but this noise is anyway expected to be subdominant to the thermorefractive and thermoelastic terms.

\subsection{Thermorefractive noise}

Ref~\cite{2004PhLA..324..345B} computed the effective temperature fluctuations for a Gaussian optical mode of $1/\mathrm{e}^2$ radius $w$ traversing a physical length $L$ through an optical solid of infinite volume.
The power spectral density of effective temperature fluctuations is 
\begin{equation}
  S_{TT}(\Omega) = \frac{k_\text{B} T^2}{\pi\kappa L} \left[\tfrac{1}{2} \rme^{+\rmi \Omega /\Omega_w} E_1(\rmi\Omega/\Omega_w) + \text{cc} \right]
\end{equation}
where $1/\Omega_w = w^2 \rho C / 4\kappa$ is the thermal relaxation time for fluctuations on a spatial scale corresponding to the beam radius $w$, and $E_1(z) = \int_z^\infty \rmd{t}\,\rme^{-t}/t$.
For $\Omega/\Omega_w \gg 1$ the term in square brackets goes to $1/(\Omega\tau_w)^2$; for $\Omega\tau_w \ll 1$, the term scales logarithmically.
The assumption of a constant $w$ is well satisfied for a cavity with $g \approx 1$.
The infinite-volume assumption becomes invalid on time scales longer than the time required for thermal fluctuations to traverse the cavity substrate; for our fused silica cavity, with height $D = \qty{25.4}{\mm}$ the corresponding Fourier frequency is $1/\tau_{D/2} = 4\kappa / \rho C (D/2)^2 = 2\pi\times\qty{3}{\mHz}$.
On the other hand, for cryogenic silicon of otherwise identical dimension, the vastly larger thermal conductivity and much smaller heat capacity yield a characteristic frequency $2\pi\times\qty{600}{\Hz}$.
Below this frequency, the infinite-volume approximation no longer holds because it includes long-wavelength thermal fluctuations that are excluded in reality from the finite-sized cavity.
The thermorefractive noise projection below this frequency is likely therefore an overestimate.
Additionally, the ringlike geometry of the optical mode has not been considered, although this is unlikely to significantly change the result.

\subsection{Thermoelastic noise}

Thermoelastic noise generates an effective displacement $x$ of the surface of a half-infinite space sensed by a Gaussian beam of width $w$ according to the formula~\cite{Cerdonio:2000ru}
  \begin{equation}
    S_{xx} = \frac{4 k_\text{B} T^2 \alpha^2 (1 + \sigma^2) w}{\sqrt{\pi} \kappa} \times J{\left(\frac{2\Omega}{\Omega_w}\right)}
  \end{equation}
where $1/\Omega_w$ is the thermal relaxation time defined previously, and~\cite{Somiya:2010zz}
  \begin{multline}
    J(u) = \real\left\{\rme^{\rmi u/2} \frac{1 - \rmi u}{u^2} \left[\erf\left(\frac{1+\rmi}{2}\sqrt{u}\right) - 1\right]\right\} \\
      + \frac{1}{u^2} - \frac{1}{\sqrt{\pi u^3}}.
  \end{multline}
Due to the cavity geometry, the displacement $x$ on a single face amounts to an optical path length shift of $n\delta L = \sqrt{2} n x$.
The uncorrelated fluctuations of four such faces therefore induce an effective frequency fluctuation $S_{\nu\nu} = 8 S_{xx}$, which is shown in \cref{fig:Si 20K nb}.
For a cryogenic silicon cavity, the thermal wavelengths below \qty{100}{\Hz} are larger than the cavity dimension and the thermoelastic fluctuations should be coherent, meaning an incoherent sum may underestimate the total thermoelastic noise amplitude by a factor of 2.
On the other hand, the computation of $S_{TT}$ assuming an infinite half-space likely overestimates the amount of low-frequency thermoelastic noise just as for the thermorefractive noise.

\subsection{Vibrational noise}

Environmentally driven vibrations at the cavity mounting points cause elastic deformation of the cavity subtrate, and hence couple laboratory acceleration into variations in the cavity optical path length.
The precise amount of noise depends in detail on the optical and mechanical geometry of the cavity and its mount, and on the laboratory environment, but we can give a rough estimate of the expected noise level and some remediation strategies here.

If the cavity is subjected to a vertical acceleration applied by a force $F = Ma = \rho H^2 L a$ to the bottom surface of area $A = L^2$, the resulting susceptibility in terms of fractional frequency shift is $(\Delta \nu/\nu_0)/a = (\Delta L / L) / a = \left. 4 \sigma \rho H^2\middle/ \sqrt{2} L E \right.$, with $E$ being the Young modulus and $\sigma$ being the Poisson ratio.
For the cavity parameters considered here, this amounts to ${} \sim \qty{e-10}{\vphantom{1}/\left({\meter\,\second^{-2}}\right)}$.
If no special geometry optimizations are pursued, this means that to reach a characteristic frequency stability of $\qty{1}{\mHz/\sqrt{\Hz}}$ requires that the cavity acceleration be kept to better than \qty{e-7}{\left({\meter/\second^2}\right)/\sqrt{\Hz}}.
Consdering that sub-hertz ground motion ranges from \qtyrange{e-9}{e-5}{\left({\meter/\second^2}\right)/\sqrt{\text{Hz}}} depending on location, time of year, and frequency band~\cite{Peterson1993}, active vibrational isolation could be needed.
Long-period seismometers can sense vibrations at the neccesary level down to millihertz frequencies~\cite{Ringler2010}, and can be integrated into active isolation platforms~\cite{Matichard:2015eva}.
Additionally, optimization of the cavity geometry and support structure can reduce the vibrational susceptibility relative to the naive case~\cite{2007PhRvA..75a1801W,2011OptL...36.3572W,2013PhRvA..87b3829L,2020RScI...91d5112W}, including in cryogenic silicon cavities where the orientation of the crystal structure is important~\cite{2012NaPho...6..687K,Zhang:2017dya}.

For the purposes of noise budgeting, we have taken the geometric mean of the Peterson low- and high-noise models for vertical ground acceleration~\cite{Peterson1993}, and used the cavity vibrational susceptibilities in \cref{tab:cavity parameters} to project this to an equivalent frequency noise; for the more advanced silicon cavity, this assumes that optimization of the mounting geometry can achieve a \qty{20}{\dB} improvement in vibrational susceptibility compared to the unoptimized case.

\begin{table*}[t]
  \centering
  \begin{tblr}{
    colspec={Q[r]Q[c]Q[l]Q[l]Q[l]},
    row{odd} = {Black!15!White},
    row{even} = {Black!10!White},
    row{1} = {bg=Black!75!White,fg=White,font=\bfseries},
  }
  Quantity & Symbol & Silica & Silicon & Unit \\
  Physical path length & $L$ &
    14.4 &
    14.4 &
    \unit{\cm} \\
  Cavity substrate side length & $H$ &
    5.08 &
    5.08 &
    \unit{\cm} \\
  Cavity substrate height & $D$ &
    2.54 &
    2.54 &
    \unit{\cm} \\
  Cavity waist size & $\sqrt{w_{\text{t}} w_{\text{s}}}$ &
    0.35 &
    0.53 &
    \unit{\mm} \\
  Temperature & $\overline{T}$ &
    293 &
    20 &
    \unit{\kelvin} \\
  Thermal conductivity & $\kappa$ &
    1.38 &
    3000 &
    \unit{\watt/(\kelvin\:\meter)} \\
  Specific heat &
    $C$ &
    730 &
    3.4 &
    \unit{\joule/(\kg\:\kelvin)} \\
  Density &
    $\rho$ &
    2200 &
    2300 &
    \unit{\kg/\meter^3} \\
  Young modulus &
    $E$ &
    72 &
    130 &
    \unit{\giga\pascal} \\
  Poisson ratio &
    $\sigma$ &
    0.17 &
    0.28 &
    {---} \\
  Refractive index &
    $n$ &
    1.45 &
    3.45 &
    {---} \\
  Thermal expansion coefficient &
    $\alpha$ &
    \num{5e-7} &
    \num{1e-8} &
    \unit{\kelvin^{-1}} \\
  Thermorefractive coefficient &
    $\beta$ &
    \num{9e-6} &
    \num{2e-6} &
    \unit{\kelvin^{-1}} \\
  Vacuum wavelength &
    $\lambda_0$ &
    1064 &
    1550 &
    \unit{\nano\meter} \\
  Cavity finesse &
    $\mathcal{F}$ &
    3000 &
    \num{20000} &
    {---} \\
  Circulating power &
    $P_\text{circ}$ &
    1.0 &
    0.01 &
    \unit{\watt} \\
  Absorbed power &
    $P_\text{abs}$ &
    \num{e-4} &
    \num{e-5} &
    \unit{\watt} \\
  Bulk loss &
    $\phi_\text{bulk}$ &
    $\num{1.2e-11}\times[f/(\qty{1}{\Hz})]^{0.77}$ &
    \num{2e-9} &
    {---} \\
  Surface loss--depth product &
    $d_\text{surf} \phi_\text{surf}$ &
    6 &
    0.5 &
    \unit{\pm} \\
  Vibrational sensitivity &
    --- &
    \num{e-10} &
    \num{e-11} &
    \unit{\vphantom{1}/(\meter\;\second^{-2})} \\
  \end{tblr}
  \caption{Table of silica and silicon cavity parameters using literature values for fused silica~\cite{Cumming:2012zz} and silicon~\cite{1959PMag....4..273F,1964PhRv..134.1058G,1978JLTP...30..621M,2010JMemS..19..229H,2012ApPhL.101d1905K,Nawrodt:2013uca,2015PhRvB..92q4113M}.
  The silica parameters represent a near-term upgrade at room temperature; the silicon version is a longer-term goal.}
  \label{tab:cavity parameters}
\end{table*}

\subsection{Shot noise and residual laser frequency noise}

Here we compute the shot noise limit for a Pound--Drever--Hall frequency locking scheme, along with an estimate of the residual frequency noise from the laser after frequency stabilization.
The shot noise limit depends on first calculating the Pound--Drever--Hall frequency discriminant that transduces frequency fluctuations relative to the cavity resonance into optical power fluctuations.

For a cavity with input and output amplitude reflectivities $r_\text{i}$ and $r_\text{e}$, and subunity internal transmission $t_\text{int}$ due to scatter or absorption losses, the cavity's field reflection transfer function is the superposition of the promptly reflected field, and a leakage of the field circulating in the cavity:
\begin{equation}
  \widetilde{r}(s) = -r_\text{i} + \frac{t_\text{i}^2 r_\text{e} t_\text{int} \rme^{-s\tau}}{1 - r_\text{i} r_\text{e} t_\text{int}\rme^{-s\tau}},
\end{equation}
where $r = r_\text{i} r_\text{e} t_\text{int}$ is the round-trip amplitude reflectivity, $\tau = n L / c$ is the round-trip travel time, and $s = \rmi\Omega$ is the Laplace-domain angular frequency corresponding to the detuning from resonance.

\begin{widetext}
To compute the PDH transfer function (discriminant), we begin with a monochromatic ingoing carrier field $E\cos(\omega_0 t)$, which has an optical-cycle-averaged power $P = A \eps_0 c \langle [E\cos(\omega_0 t)]^2\rangle_{\omega_0} = A \eps_0 c E^2/2$, where $A$ is an effective beam area and $\eps_0$ is the vacuum permittivity.
When a sinusoidal phase modulation $\Gamma\sin{(\Omega_\text{mod}t)}$ is applied to this field, to leading order in the modulation index $\Gamma$ the resulting field is $\tfrac{1}{2} E \left[ J_0(\Gamma) + \rmi J_1(\Gamma) \rme^{-\rmi\Omega_\text{mod} t} -\rmi J_1(\Gamma) \rme^{+\rmi\Omega_\text{mod} t} \right] \rme^{-\rmi\omega_0 t} + \text{cc}$, with $J_1$ being the first order Bessel function of the first kind.
Suppose the carrier field couples into the cavity with power fraction (visibility) $V$.
The sidebands (at $\omega_0\pm\Omega_\text{mod}$) are promptly reflected.
Upon reflection from the cavity, one therefore has a field
\begin{equation}
  \tfrac{1}{2} E_\text{refl} \rme^{-\rmi\omega_0 t} + \text{cc} = \tfrac{1}{2} E \left[ J_0(\Gamma) \left(\sqrt{V}\, \widetilde{r}(\Omega) + \sqrt{1 - V}\, \widetilde{r}_\text{prompt}\right)+ 2\rmi J_1(\Gamma) \sin\left(\Omega_\text{mod} t\right) \right] \rme^{-\rmi\omega_0 t} + \text{cc},
\end{equation}
where $\Omega \equiv \omega_0 - \omega_\text{res}$ is the detuning of the laser relative to the cavity eigenfrequency.
We have assumed $r(\Omega_\text{mod})$ is constant and common to both the upper and lower PDH sidebands.
The optical-cycle-averaged power on reflection is
\begin{equation}
  P_\text{refl}(t) = A \eps_0 c \left\langle \left|\tfrac{1}{2} E_\text{refl} \rme^{-\rmi\omega_0 t} + \text{cc}\right|^2\right\rangle_{\omega_0} = A \eps_0 c \frac{|E_\text{refl}|^2}{2},
\end{equation}
which contains a term proportional to $\sin(\Omega_\text{mod} t)$.
When the reflected optical-cycle-averaged power is demodulated at $\Omega_\text{mod}$, the resulting rf amplitude is (neglecting factors such as mixer inefficiencies)
\begin{equation}
  \langle P_\text{refl} \rangle_{\Omega_\text{mod}} = -\frac{\rmi}{2} A \eps_0 c E^2 J_0(\Gamma) J_1(\Gamma) r^*(\Omega_\text{mod}) \left[ \sqrt{V} \widetilde{r}(\Omega) + \sqrt{1 - V} r_\text{prompt}\right] + \text{cc}.
\end{equation}
\end{widetext}
If $r(\Omega_\text{mod})$ and $r_\text{prompt}$ are further assumed to be real, the resulting demodulated rf power is
\begin{equation}
  \langle P_\text{refl} \rangle_{\Omega_\text{mod}} = 2 P J_0(\Gamma) J_1(\Gamma) \widetilde{r}(\Omega_\text{mod}) \sqrt{V} \imag{\widetilde{r}(\Omega)}.
\end{equation}
To leading order in $\Omega$,
\begin{equation}
  \imag{\widetilde{r}(\Omega)} = \left(\frac{t_\text{i}}{1 - r_\text{i} r_\text{e} t_\text{int}}\right)^2 r_\text{e} t_\text{int}\, \Omega\, \tau,
  \label{eq:Im PDH}
\end{equation}
where $[t_\text{i}/(1-r_\text{i} r_\text{e} t_\text{int})]^2 \equiv \mathcal{G}_\text{cav}$ is the power gain of the cavity, equal to the ratio $P_\text{circ} / [J_0(\Gamma)^2 V P]$ of intracavity circulating power to the ingoing carrier power that is available to couple into the cavity.\footnote{%
  In the case when the input and output couplings are equal ($t_\text{i}^2 = t_\text{e}^2$) and when optical losses at the input and output coupling interfaces are negligible ($r_\text{i}^2 + t_\text{i}^2 = r_\text{e}^2 + t_\text{e}^2 = 1$), and the internal losses are much smaller than the input and output couplings ($1 - t_\text{int}^2 \ll t_\text{i,e}^2$), then the gain and finesse are related by $\mathcal{G}_\text{cav} = \mathcal{F}/\pi$.
}
If $r_\text{e}$ and $t_\text{int}$ are close to 1 in \cref{eq:Im PDH}, and $\widetilde{r}(\Omega_\text{mod}) \approx 1$ as well, then the PDH discriminant, in demodulated W/Hz of detuning, is $\mathcal{D} = 4\pi P \sqrt{V} J_0(\Gamma) J_1(\Gamma) \mathcal{G}_\text{cav} n L /c$ within the cavity bandwidth.

The shot noise on reflection, accounting for cyclostationary effects, has power spectral density~\cite{Niebauer:1991krz,McKenzieThesis,Chalermsongsak:2014aua}
\begin{equation}
  S_{PP} = 2h\nu_0 P \times \left\{\left[(1 - V) + V |\widetilde{r}(0)|^2 \right] J_0(\Gamma)^2 + 3 J_1(\Gamma)^2 \right\},
\end{equation}
which again assumed all prompt reflections are simply unity.
In the limit of perfect visibility ($V = 1$) and critical input coupling ($\widetilde{r}(0) = 0$), the frequency-referred shot noise limit (using the previous expression for the discriminant $\mathcal{D}$) is
\begin{equation}
  S_{\nu\nu}(\Omega) = \frac{c}{\pi n L} \sqrt{\frac{3h\nu_0}{8 \mathcal{G}_\text{cav} P_\text{circ}}},
\end{equation}
which is valid for Fourier frequencies within the cavity bandwidth.

\begin{figure}[t]
  \centering
  \includegraphics[width=\columnwidth]{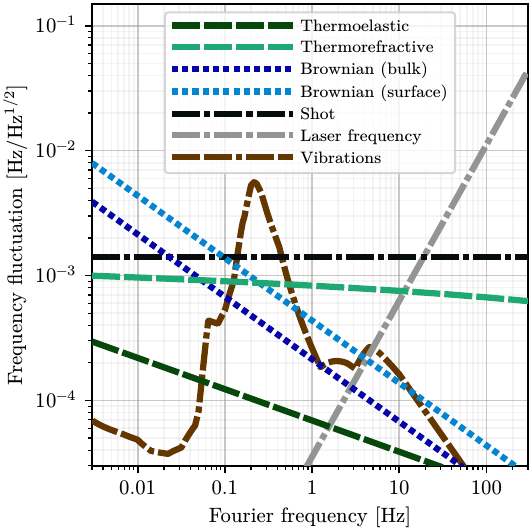}
  \caption{Noise budget of the absolute frequency stability of a cryogenic crystalline silicon cavity operated at \qty{20}{\kelvin} and pumped with \qty{1550}{\nm} light.}
  \label{fig:Si 20K nb}
\end{figure}

The light may also contain residual frequency noise contributions from the laser due to the finite suppression of the frequency servo.
For a \qty{1064}{\nm} Nd:YAG nonplanar ring oscillator, the freerunning noise is typically $\left(\qty{100}{\Hz/\sqrt{\Hz}}\right)\times(\qty{100}{\Hz}/f)$.
For \qty{1550}{\nm}, the frequency noise was assumed to be $\left(\qty{1}{\kHz/\sqrt{\Hz}}\right)\times\sqrt{(\qty{100}{\Hz})/f}$, which is representative of a number of doped-fiber or semiconductor-based lasers~\cite{Meylahn:2021dwh}.
A frequency servo with a bandwidth of \qty{100}{\kHz} and several stages of integration near \qty{10}{\kHz} was assumed to project the residual laser frequency noise in \cref{fig:Si 20K nb}, but further suppression could be achieved with a more aggressive servo controller design or by prestabilizing the laser to another cavity.

\subsection{Photothermal fluctuations}

Temperature fluctuations in excess of the thermodynamic limit have the potential to degrade the cavity frequency stability.
This is expected to be a more pressing issue for a cryogenic cavity, where the frequency stability target is more stringent.
A full numerical analysis of the cavity stability depends on the exact implementation of the thermal shielding and cryogenic configuration, which is premature; here we give an analytical calculation that argues for the basic feasibility of such a cryogenic TIR cavity.

The overall behavior and scale of the thermal problem can be estimated in a simplified cylindrical geometry, where a linear Gaussian beam mode of $1/\rme^2$ radius $w$ is oriented along the $z$ axis, the cavity volume extends from $z = 0$ to $z = L$, and radially from $r = 0$ to $r = D/2$.
The cavity boundary is allowed to radiate to a shield at a temperature much colder than the temperature $\overline{T} = \qty{20}{\kelvin}$ at the cavity surface.
The radiative area is comparable to $2H^2$, the combined area of the top and bottom cavity surfaces.
Before discussing fluctuations, several constraints arise from a consideration of the static temperature.
The maximum power that can be radiated from the cavity is $\sigma_\text{SB} \overline{T}^4 \times 2 H^2 \approx \qty{50}{\micro\watt}$; if the cavity absorption amounts to an average of \qty{5}{ppm/\cm} per unit path length, then choosing a circulating cavity power \qty{0.3}{\watt} translates to a static absorption $\overline{P}_\text{abs} = \qty{20}{\micro\watt}$, which is taken as the nominal value for subsequent calculations.

In this geometry, the heat flow is independent of $z$; in the frequency domain, the time-varying part of the heat equation reads
\begin{equation}
  \rmi\rho C \Omega T(r,\Omega) = Q(r,\Omega) - \kappa \nabla^2 T(r,\Omega),
\end{equation}
defined on the interval $0 \le r \le D/2$; at $r = D/2$, we assume a radiative boundary condition $-\kappa \partial_r T(D/2,\Omega) = 4 \sigma \overline{T}(D/2)^3 T(D/2,\Omega)$, where $\overline{T}(D/2)$ is the static temperature at the boundary.
The problem is solved by expanding both $T(r,\Omega)$ and $Q(r,\Omega)$ in Dini series, summing over Bessel functions $J_0(\lambda_n D/2)$, where $\lambda_n$ solves $\kappa\lambda_n J_1(\lambda_n D/2) = 4 \eps_\text{bd}\eps_\text{shield} \sigma_\text{SB} \overline{T}(D/2)^3 J_0(\lambda_n D/2)$.
The Dini series for the injected heat is
\begin{equation}
  Q(r,\Omega) = \sum_{n=1}^\infty Q_n J_0(\lambda_n r)
\end{equation}
with
\begin{equation}
  Q_n = \frac{1}{C_n} \int\limits_0^{D/2}\hspace{-1ex}\rmd r\, r\, J_0(\lambda_n r) Q(r,\Omega)
\end{equation}
and
\begin{equation}
  C_n = \int\limits_0^{D/2}\hspace{-1ex}\rmd r\, r\, J_0(\lambda_n r)^2 = \frac{D^2}{8}\left[J_0(\lambda_n D/2)^2 + J_1(\lambda_n D/2)^2\right].
\end{equation}
Similarly, we can expand
\begin{equation}
  T(r,\Omega) = \sum_{n=1}^\infty T_n J_0(\lambda_n r)
\end{equation}
and the linearity of the heat equation implies
\begin{equation}
  T_n = \frac{Q_n}{\kappa \lambda_n^2 + \rmi\rho C \Omega}.
\end{equation}
Because $w \ll D/2$, the coefficients $Q_n$ can be approximated:
\begin{equation}
  Q_n \xrightarrow{w\ll D/2} \frac{P}{2\pi L C_n} \exp\left(-\frac{\lambda_n^2 w^2}{8}\right).
\end{equation}

For the photothermal transfer function, we define
\begin{equation}
  \hat{T}(\Omega) = \frac{4}{w^2} \int\limits_0^\infty \rmd r\,r\,T(r,\Omega) \rme^{-2 r^2/w^2}
\end{equation}
which in the $w \ll D/2$ limit has
\begin{equation}
  \hat{T}(\Omega) \xrightarrow{w \ll D/2} \frac{P_\text{abs}(\Omega)}{2\pi L} \sum_{n=1}^\infty \frac{(1/C_n)\rme^{-\lambda_n^2 w^2 / 4}}{\kappa \lambda_n^2 + \rmi \rho C \Omega}.
\end{equation}
In the infinite-volume limit, the corresponding expression would instead be $\hat{T}(\Omega) = [P_\text{abs}(\Omega)/4\pi\kappa L] \rme^{\rmi\Omega/\Omega_w} E_1(\rmi\Omega/\Omega_w)$; the discrepancy between the finite- and infinite-volume cases is only logarithmic and emerges for $\Omega \ll \rho C (D/2)^2/4\kappa$.
For the silicon geometry considered, the low-frequency finite-volume transfer function has a typical scale $\hat{T}(0)/P_\text{abs}(0) \sim \qty{e-3}{\kelvin/\watt}$.
The amplitude spectral density of absorbed optical power $\sqrt{2 h \nu \overline{P}}$ implies a thermorefractively-coupled photothermal noise less than $\qty{e-6}{\Hz/\sqrt{\Hz}}$, which is much smaller than the fundamentally driven thermal noises.

\end{document}